\documentclass[sigconf]{acmart}

\makeatletter
\def\mdseries@tt{m}
\makeatother
\usepackage[plain]{fancyref}
\usepackage[draft=true]{minted}
\usepackage{color}
\usepackage{hyperref}
\hypersetup{
    colorlinks=true,
    linkcolor=blue,
    filecolor=red,      
    urlcolor=magenta,
    breaklinks=true,
}
\usepackage{breakurl} 

\begin{document}

\sloppy 

\newcommand{\ie}{i.e.,\,}
\newcommand{\eg}{e.g.,\,}
\newcommand{\etal}{et al.}
\newcommand{\confirm}[1]{{\color{red} \textbf{Needs confirmation} {#1}}}
\newcommand{\doweneedthis}[1]{{\color{orange} \textbf{Do we need this?} {#1}}}
\newcommand{\note}[1]{{\color{purple} \textbf{Note} {#1}}}
\newcommand{\editFromHere}{{\color{blue} \textbf{Continue Editing From Here}}}
\newcommand{\confirmed}{{\color{green} confirmed}}

\title{Ghera: A Repository of Android App Vulnerability Benchmarks}

\author{Joydeep Mitra}
\affiliation{
  \institution{Kansas State University}
  \country{USA} 
}
\email{joydeep@k-state.edu}

\author{Venkatesh-Prasad Ranganath}
\affiliation{
  \institution{Kansas State University}
  \country{USA} 
}
\email{rvprasad@k-state.edu}

\renewcommand{\shortauthors}{Mitra and Ranganath}

\begin{abstract}
Security of mobile apps affects the security of their users.  This has fueled the development of techniques to automatically detect vulnerabilities in mobile apps and help developers secure their apps; specifically, in the context of Android platform due to openness and ubiquitousness of the platform.  Despite a slew of research efforts in this space, there is no comprehensive repository of up-to-date and lean benchmarks that contain most of the known Android app vulnerabilities and, consequently, can be used to rigorously evaluate both existing and new vulnerability detection techniques and help developers learn about Android app vulnerabilities.  In this paper, we describe \emph{Ghera}, an open source repository of benchmarks that capture 25 known vulnerabilities in Android apps (as pairs of exploited/benign and exploiting/malicious apps).  We also present desirable characteristics of vulnerability benchmarks and repositories that we uncovered while creating Ghera.
\end{abstract}

\keywords{Security, Collection, Benchmarking, Mobile, Applications}

\copyrightyear{2017} 
\acmYear{2017} 
\setcopyright{acmlicensed}
\acmConference{PROMISE}{November 8, 2017}{Toronto, Canada}
\acmPrice{15.00}
\acmDOI{10.1145/3127005.3127010}
\acmISBN{978-1-4503-5305-2/17/11}

\maketitle

\section{Introduction}
\label{sec:introduction}

\subsection{Motivation}
\label{sec:intro-motivation}

With increased use of mobile devices (including mobile phones), mobile apps play a pivotal role in our lives.  They have access to and use huge amounts of sensitive and personal data to enable various services  banking, shopping, social networking, and even two-step authorization.  Hence, security of mobile devices and apps is crucial to guarantee the security and safety of their users.  

This observation is further amplified on the Android platform as it is widely adopted by both consumers (users) and developers -- Android has captured 88\% of the global market share as of third quarter of 2016 \cite{MarketShare1:URL} and Google Play, the official store for Android apps, has 2.8 million apps as of March 2017 \cite{GooglePlayAppCount:URL}.

One approach to secure Android devices is to keep malicious apps out of Android devices.  This approach is supported by numerous tools and techniques that detect malicious behaviors in apps \cite{Sadeghi:TR2016,Sufatrio:CSUR15,Bouncer:URL,Zhou:IEEESSP12,Zhou:NDSS12}.  However, these techniques cannot detect every malicious behavior.  Further, malicious behaviors often depend on vulnerabilities in the Android platform and the apps executed on the platform.  

Consequently, a complementary approach is to \emph{secure (harden) Android apps}.  This approach is also supported by tools and techniques to detect known vulnerabilities in apps (referred to as \emph{app vulnerabilities}) that stem from incorrect use of or bugs in Android framework\footnote{Android framework is a set of APIs available to Android apps to interact with the Android platform.} and can be used to carry out malicious actions such as data theft.\footnote{This paper does not explore the application of this approach to harden the Android platform.}  

Like all automation-based approaches, this approach is useful only if developers can trust the verdicts of the employed tools and techniques.  Such trust can be established only by rigorous and reproducible evaluations of tools and techniques.  By \emph{rigorous}, we mean the verdict of a technique can be verified to be true and to be caused only by the reasons/explanations provided by the technique (within the context considered in the evaluation).  By \emph{reproducible}, we mean evaluations can be repeated to verify the reproduction of (same) results.  

Further, if such evaluations are based on \emph{a common baseline -- a set of benchmarks (apps)\footnote{A benchmark is a standard or point of reference used for evaluation/comparison.  So, in this paper, we refer to an app that embodies a specific vulnerability X (along with an app that exploits X) as a benchmark for X.} containing specific vulnerabilities}, then their results will be fair and comparable.  By \emph{fair}, we mean evaluations will not favor any specific technique.  By \emph{comparable}, we mean the results from evaluations of different techniques can be compared (as the evaluations are controlled for subjects).  Consequently, developers can use the results from such evaluations to easily compare and select techniques based on aspects such as efficacy, efficiency/scale, and ease-of-use.  Also, a common baseline can simplify comparison of future techniques with existing techniques and help make robust claims about future techniques.

Another approach to secure Android apps is to \emph{educate developers about securing apps}.  This approach is enabled by the extensive official documentation provided by Google about security in Android and the best practices for security and privacy in Android apps \cite{AndroidSecurity:URL,AndroidSecurityBestPractices:URL}.  It is also enabled by documentation available in the form of white papers, guides, and books about how to secure Android apps \cite{McAffee:Rudrapp15,JSSEC:Book,Elenkov:Book}.  Most of these resources provide code snippets (at times, even apps) that showcase good practices to secure apps.

Despite the availability of such resources, Android apps still have vulnerabilities \cite{CVE:URL,Reaves:USENIX15}.  The presence of known vulnerabilities in apps can only stem from developer's lack of knowledge about known vulnerabilities, benefits of securing apps, or how security can easily go awry.  Like bad code examples are interspersed between good code examples in programming books to highlight good practices via contrasting, a way to help address such lack of knowledge is to \emph{offer apps that showcase known vulnerabilities and demonstrate how security can easily go awry.}  Such a repository of vulnerable apps can also serve as an channel to learn about known vulnerabilities in Android apps.

Finally, while there are numerous efforts to develop tools and techniques to secure Android apps, there is no single benchmark repository that captures most of the known vulnerabilities in Android apps in a technique/tool agnostic manner.

\subsection{Contributions}
\label{sec:intro-contributions}

Motivated by above observations and the need of an ongoing effort to assess existing vulnerability detection tools for Android, we created \emph{Ghera},
an open source repository of vulnerability benchmarks\footnote{We refer to a benchmark containing at least one vulnerability as \emph{a vulnerability benchmark}.}.  Currently, Ghera comprises 25 benchmarks.  Each benchmark is a pair of exploited/benign and exploiting/malicious apps that captures a unique known vulnerability in Android apps.  The benchmarks span four areas of Android framework: \emph{Inter Component Communication (ICC), Storage, System,} and \emph{Web}.  Ghera is open sourced under BSD 3-clause license and publicly available at \url{http://bitbucket.org/secure-it-i/android-app-vulnerability-benchmarks}.

While creating these benchmarks, we found very little guidance in the literature regarding how to create good benchmarks or the desirable characteristics of good benchmarks.  So, after creating the benchmarks, we did a retrospection and identified various desirable characteristics of vulnerability benchmarks, which are applicable to benchmarks in general.  

In this paper, we present Ghera in detail along with the desirable characteristics of vulnerability benchmarks.

The remainder of the paper is organized as follows. \Fref{sec:characteristics} lists desirable characteristics of vulnerability benchmarks. It also explores related work in the light of these characteristics. \Fref{sec:repository} describes Ghera in terms of its design choices, structure and content, characteristics, and limitations. \Fref{sec:future-work} lists future possibilities with Ghera.  \Fref{sec:catalog} catalogs the vulnerabilities captured in Ghera.

\section{Desirable Characteristics of Vulnerability Benchmarks} 
\label{sec:characteristics}

\subsection{Context}
\label{sec:cha-context}

When we started an effort to assess various vulnerability detection techniques, we searched for suites of tests or benchmarks to fairly evaluate the techniques.  We did not find such suites.  Most existing efforts used applications from app stores, \eg Google Play.  While few efforts constructed small dedicated example apps, almost all of the example apps were geared towards demonstrating the associated techniques and did not explicitly capture known vulnerabilities, \eg DroidBench.  Hence, we took a detour in our effort to \emph{collect and catalog known Android app vulnerabilities in an informative and comprehensive repository to enable rigorous and reproducible evaluation of vulnerability detection tools and techniques}.

As we started collecting vulnerabilities, we searched for guidance to create good benchmarks.  To our surprise, while there were numerous benchmarks, there was very little information about how to design good benchmarks or even characteristics of a good benchmark.  Hence, after we collected vulnerabilities, we did a retrospection to identify characteristics that we considered while collecting vulnerabilities along with the reasons why we considered them.

\subsection{Vulnerability Benchmark Characteristics}
\label{sec:cha-benchmarks}

Here are the characteristics identified during our retrospection.  Besides describing the characteristics, we also describe how they were influenced by existing efforts in the space of detecting vulnerabilities in Android apps.

\subsubsection{Tool and Technique Agnostic} \emph{The benchmark is agnostic to tools and techniques and how they detect vulnerabilities.}  This characteristic enables the use of benchmarks for fair evaluation and comparison of tools and techniques.

We uncovered this characteristic when we explored \textit{DroidBench}\footnote{https://github.com/secure-software-engineering/DroidBench}, one of the first benchmark suites created in the context of efforts focused on detecting vulnerabilities in Android apps \cite{Arzt:PLDI14}.  DroidBench is tailored to evaluate the effectiveness of taint-analysis\footnote{Taint analysis identifies parts of the program affected (tainted) by specific data sources such as user input.} tools to detect information leaks in Android apps.  So, the benchmarks are geared towards testing the influence of various program structures and various aspects of static analysis (\eg field sensitivity, trade-offs in access-path lengths) on the effectiveness of taint analysis to detect information leaks (vulnerabilities).  Further, it is unclear if program structures considered in the benchmarks reflect program structures that enable vulnerabilities in real world apps or program structures that push the limits of static analysis techniques.  Hence, these benchmarks are not tool and technique agnostic.

In contrast, repositories such as AndroZoo\footnote{https://androzoo.uni.lu/} \cite{Allix:MSR16} and PlayDrone\footnote{https://github.com/nviennot/playdrone} \cite{Viennot:SIGMETRICS14} provide real world Android apps available in various app stores.  Further, the selection of apps is independent of their intended use by any specific tool or technique.  Hence, any vulnerable apps (benchmarks) in these repositories are tool and technique agnostic.

\subsubsection{Authentic} \emph{If the benchmark claims to contain vulnerability X, then it truly contains vulnerability X.}  This characteristic enables benchmarks to serve as ground truths when evaluating accuracy of tools and techniques.  Consequently, the comparison of tools and techniques is simplified.

To appreciate this characteristic, consider the evaluation of MalloDroid, a tool to detect the possibility of Man-in-the-Middle (MitM) attacks on Android apps due to improper TLS/SSL certificate validation \cite{Fahl:CCS12}.  To evaluate the accuracy of MalloDroid, 13,500 apps were analyzed using Mallodroid and 8\% of them were flagged as potentially vulnerable.  However, since the potential vulnerability does not imply the vulnerability can be exploited to steal information, 266 apps from Google Play were selected based on app categories that are likely to be most affected by detected vulnerabilities.  Of these apps, 100 of the most popular apps were manually audited and 41 of them were found to have exploitable vulnerabilities.  Further, not all aspects of apps were exercised during the evaluation.  This evaluation could have been simpler, easier, more complete, and more accurate using authentic vulnerable benchmarks. 


In terms of comparing tools and techniques, when EdgeMiner \cite{Cao:NDSS15} was compared to FlowDroid \cite{Arzt:PLDI14}, 9 apps were flagged with vulnerabilities by EdgeMiner but not by FlowDroid.  To verify this outcome, the 9 flagged apps were analyzed using TaintDroid, yet another tool, and 4 apps were flagged by TaintDroid. So, it was concluded that EdgeMiner did better than FlowDroid in 4 cases.  With authentic benchmarks, the comparison would have been simpler and extensive.

Of the various repositories we explored, AndroZoo provides some evidence of malicious behavior (exploits) in apps in the form of verdicts from running analysis tools on apps.  However, no evidence is provided about presence of vulnerabilities.  Hence, these benchmarks cannot be used authentic vulnerability benchmarks.  In contrast, the benchmarks in DroidBench are authentic as they are seeded with vulnerability during construction.

\subsubsection{Feature Specific} \emph{If the benchmark uses only features F of a framework to create vulnerability X, then the benchmark does not contain other features of the framework that can be used to create X.}  This characteristic helps evaluate if tools and techniques can detect vulnerabilities that stem only due to specific reasons (features).  In other words, it helps assess if and how the cause of a vulnerability affects the ability of tools and techniques to detect the vulnerability. Often, this could translate into being able to verify the explanations provided by a tool when it detects X.

As discussed above, EdgeMiner detected 4 more vulnerable apps than FlowDroid.  However, there was no explanation for the better performance of EdgeMiner in terms of features (causes) that EdgeMiner handled better than FlowDroid.  Such explanations could have been easily uncovered with feature specific benchmarks.

Often, real world apps serving as benchmarks (as in case of AndroZoo and PlayDrone) lack this characteristic as the causes of app vulnerabilities in them are most likely unknown to the public.  In contrast, benchmarks in repositories such as DroidBench exhibit this characteristic as they are feature specific by construction.

\subsubsection{Contextual} \emph{The benchmark capturing vulnerability X in a context C is distinct from benchmarks capturing X in other contexts.}  This characteristic helps evaluate the efficacy of tools and techniques to detect vulnerabilities in specific contexts, \eg real world scale, experimentation, use of specific libraries. 

To understand this characteristic, consider the size of benchmarks.  While evaluating a tool (or a technique) to detect vulnerabilities, we can first evaluate it on \emph{lean} benchmarks -- \emph{A benchmark containing vulnerability X that can be created by any of the features F of a framework is lean if it makes minimal use of features of the framework not in F} -- .  Due to leanness, such benchmarks can speed up the evaluation process and enable easy comprehension when debugging/understanding the behavior of the tool.  Further, such benchmarks can be used as tests while building tools.  When in doubt, authenticity of such benchmarks can be easily verified with manual effort. After such evaluations, we can consider \emph{fat (non-lean)} benchmarks to evaluate how the tool performs (scales) on larger inputs.  Hence, with contextual benchmarks, efficacy evaluations can be more focused and streamlined.

In our explorations, we found tool evaluations that use both lean and fat benchmarks \cite{Arzt:PLDI14,Wei:CCS14} and tool evaluations that use only fat benchmarks \cite{Fahl:CCS12,Octeau:SEC13,Sadeghi:ICSE15}\footnote{While lean benchmarks may have been used, such uses were not reported.}.  In comparison, establishing veracity of tools was easier in former tool evaluations.

As for repositories, DroidBench offers custom apps as lean benchmarks while AndroZoo and PlayDrone offer real world apps as fat benchmarks.

Like size, connectivity, resource availability, complexity, number of vulnerabilities in a benchmark, and number of kinds of vulnerabilities in a benchmark can be considered to identify useful contexts when designing benchmarks.

\subsubsection{Ready-to-Use} \emph{The benchmark is composed of artifacts that can be used as is to reproduce the vulnerability.}  This characteristic precludes the influence of external factors (\eg interpretation of instructions, developer skill) in realizing a benchmark, \eg starting from its textual description or skeletal form.  Hence, it enables fair evaluation and comparison of tools and techniques.

DroidBench, AndroZoo, and PlayDrone repositories provide benchmarks as ready-to-use APKs (Android app bundles).

In comparison, SEI provides a set of guidelines for development of secure Android apps \cite{SEIRules:URL}.  The descriptions of many guidelines are accompanied by illustrative good and bad code snippets.  While the code snippets are certainly helpful, they are not ready-to-use in the above sense.  This is also true of many security related code snippets available as part of Android documentation.

\subsubsection{Easy-to-Use} \emph{The benchmark is easy to set up and reproduce the vulnerability.}  Benchmarks with this characteristic help expedite evaluations.  Consequently, this characteristic can help usher wider adoption of the benchmarks.  This characteristic is desirable of benchmarks that require some assembling, \eg build binaries from source, extensive set up after installation.

As with ready-to-use characteristic, DroidBench, AndroZoo, and PlayDrone cater binary benchmarks that are easy to install and conduct evaluations.  The source form of benchmarks provided by DroidBench also have this characteristic as they contain Eclipse project files required to build them.

\subsubsection{Version Specific} \emph{The benchmark is associated only with the versions of the framework in which the contained vulnerability can be reproduced.}  With this characteristic, benchmarks can be chosen for an evaluation based on the version of the framework being used in the evaluation.  Hence, it helps evaluations to be version specific.

To appreciate this characteristic, consider a vulnerability that was affected by the rapid evolution of Android framework -- evolved from level (version) 1 thru 25 from 2008 to 2017.  In 2011, when the latest version of Android was 4.0.4, Chin \etal\, revealed that a background service in an Android app could be hijacked by a malicious app installed on the device if the service allowed clients to start the service via an implicit intent \cite{Chin:Mobisys11}.  Starting with Android 7.0 in 2016, the vulnerability was invalidated as starting of services via implicit intent were prohibited.

Now, suppose an evaluation uses a benchmark that captures this vulnerability but is not version specific.  This can lead to two undesirable situations.  In the first situation, a tool targeting Android 7 apps will be flagged as incorrect  when it (correctly) fails to detect this vulnerability.  In the second situation, a tool that detects the vulnerability will be flagged as correct but it will incorrectly detect non-existent vulnerabilities in Android 7 apps.  Both these situations can be avoided by using version specific benchmarks.

In terms of maintenance and keeping benchmarks current, version specific benchmarks can be easily updated to reflect any changes (such as end of life support) to associated versions of Android framework/platform.  This will prevent accidental evaluations of benchmarks with unsupported versions of Android.

The benchmarks in AndroZoo, PlayDrone, and DroidBench are not version specific as these benchmarks have no information about compatible versions of Android framework/platform.  Consequently, DroidBench can enable a situation similar to that one described above.  Many Droidbench benchmarks use log files as information sink.  Such use was valid prior to Android 4.1 as every app on a phone could read log files of any apps on the phone.  However, since Android 4.1, apps can read only their log files.  Hence, these benchmarks can lead to incorrect evaluations depending on Android version being considered.

\subsubsection{Well Documented} \emph{The benchmark is accompanied by relevant documentation.}  Such documentation should contain description of the contained vulnerability and the features used to create the vulnerability.  It should also mention the target (compatible) versions of the framework/platform and provide instructions to both surface the vulnerability and exploit the vulnerability.  When possible, the source code of the benchmark should be included as part of the documentation.  This characteristic obviously helps expedite evaluations that use the benchmarks and contributes to ease of use of benchmarks.  With source code, it can help developers understand the vulnerability.

The benchmarks provided by DroidBench are in some ways well documented as they contain source code along with binaries and there is brief documentation on the web site and in the source code about captured vulnerabilities.  This is not the case with benchmarks catered by AndroZoo and PlayDrone.

\subsubsection{Dual} \emph{The benchmark contains both the vulnerability and a corresponding exploit (dual).}  This characteristic simplifies evaluations that depend on exercising the vulnerability, \eg dynamic analysis.  It allows benchmarks to be used to demonstrate vulnerabilities and even evaluate exploits.  Benchmarks with this characteristic can help developers understand the vulnerability; specifically, when the source code is available.  Also, duality helps verify the authenticity of benchmarks.

In our exploration, we did not find any dual benchmarks.

\subsection{Vulnerability Benchmark Repository Characteristics}
\label{sec:cha-repos}

Similar to the above benchmark characteristics, here are two desirable characteristics of vulnerability benchmark repositories.

\subsubsection{Open} \emph{The benchmark repository should open to the community both in terms of consumption and contribution.}  The benchmarks should be available with minimal restrictions (\eg permissive licence) and preferably at no or very low cost to the community.  The repository should have a well-defined yet accessible process for the community to contribute new benchmarks.  This characteristic helps with reproducibility of results and community wide consolidation of benchmarks.  The latter effect reduces duplication efforts in the community.

In this regard, DroidBench is more open than AndroZoo and PlayDrone repositories.  DroidBench is hosted as a public repository on GitHub and it welcomes contributions.  PlayDrone is hosted as multiple public archives on Internet Archive with no explicit guidance for contributions.  AndroZoo is hosted as a web service that can be accessed only by approved users.  This is most likely to manage and track access to a large corpus of data in AndroZoo.  Like PlayDrone, there is no explicit guidance for contributions. This may be due to how AndroZoo is populated -- with real world apps collected from different app stores.

\subsubsection{Comprehensive} \emph{The benchmark repository should have benchmarks that account for (almost) all known vulnerabilities of the target framework/platform.}  This characteristic simplifies evaluations as they can rely on a single repository (or very few repositories) to consider all vulnerabilities.  Further, evaluations can be more thorough as they can consider most of the known vulnerabilities.

In our explorations, we did not find a single repository that covered almost all vulnerabilities in Android apps.  While DroidBench does a pretty good job covering information leak vulnerabilities stemming from ICC, it does not cover information leaks due to other reasons such as misuse of WebView component \cite{Luo:ACSAC11}.  In terms of evaluation, Reaves \etal\,\cite{Reaves:CSUR16} used DroidBench along with 6 mobile money apps \cite{Reaves:USENIX15} and 10 most widely used financial apps in Google Play to evaluate Android security analysis tools.  While DroidBench and 6 mobile money apps had certain known vulnerabilities, they did not cover all kinds of vulnerabilities.  With a comprehensive repository, this evaluation could have been simpler and more thorough.

\subsection{Discussion}
\label{sec:cha-general}

\paragraph{Reality Check} Of the above characteristics, some are easier to achieve than others.  For example, creating an open repository is easier than creating a comprehensive repository.  Similarly, creating a tool agnostic benchmark is easier than creating an authentic benchmark.  Further, while it is desirable for benchmarks/repositories to have all of the characteristics, it is hard to achieve them all as we have seen in examples.  Nevertheless, given the benefits of these characteristics, we believe that the community should strive to create benchmarks/repositories with these characteristics.

\paragraph{Wider Relevance} While we uncovered and described these characteristics in the context of vulnerabilities, we believe they apply to benchmarks in general; say, in other contexts such as performance.  For example, in the context of performance benchmarking, \emph{agnostic} characteristic can be restated as \emph{The benchmark is agnostic to techniques and how they achieve performance}.

\section{Ghera}
\label{sec:repository}

We created Ghera, an Android app vulnerability benchmarks repository, because we needed vulnerable benchmarks to evaluate existing tools that aid in the development of secure Android apps.  We initially explored existing repositories like DroidBench, AndroZoo, and PlayDrone.  Except for DroidBench, there was little to no information about the presence and kind of vulnerabilities in apps from these repositories.  In case of DroidBench, the benchmarks/apps were specific to information flow-based vulnerabilities.

So, our goals while creating Ghera were to ensure the benchmarks contained unique Android app vulnerabilities and were tool agnostic, easy to use, and well documented with source code.  In addition, we wanted the repository to be open and as comprehensive as possible.

\subsection{Design Choices}
\label{sec:repo-design-choices}

Given our goals for the repository, we decided to group vulnerabilities based on the features (capabilities) that cause the vulnerabilities.  To decide on the features that we wanted to explore, we looked at features commonly used by Android apps and discussed in various Android security related resources \cite{Chin:Mobisys11,Fahl:CCS12,McAffee:Rudrapp15,AndroidSecTips:URL}.  Almost all Android apps use one or more of the following capabilities: 

\begin{itemize}
  \item Communicate with components in apps installed on the device.
  \item Store data on the device.
  \item Interact with the Android platform.
  \item Use web services.
\end{itemize}

Based on these capabilities, for the initial version of the repository, we identified \emph{Inter Component Communication (ICC), Storage, System,} and \emph{Web} as the categories of vulnerabilities.

In each category, we identified APIs pertinent to that category and studied them.  To determine potential vulnerabilities stemming from an API, we primarily explored prior research efforts, Stack Overflow discussions, Android documentation, and the source code in the Android Open Source Project (AOSP).\footnote{We explored public facing lists of vulnerabilities and exploits maintained by organizations such as Mitre \cite{CVE:URL}.  While these lists are useful to understand different kinds of common vulnerabilities on a platform or a framework, they often lack details about app vulnerabilities such as primary causes of a vulnerability or how to reproduce a vulnerability.  Also, many of these vulnerabilities do not have accompanying known exploits.  So, to quickly bootstrap the benchmark repository, we did not consider such lists.  However, since these lists serve as excellent starting points to create benchmarks for app vulnerabilities, we are exploring them to expand the repository.}  When we uncovered a potential vulnerability X related to the API, we developed an app M with the vulnerability X along with an app N to exploit vulnerability X in app M.  We then verified the vulnerability by executing apps M and N and checking if the vulnerability was indeed exploited.  This verification was carried out on Android versions 4.4 thru 7.1.

We decided to name each benchmark based on the feature causing the vulnerability captured by the benchmark and the exploit used to confirm the vulnerability.  So, a benchmark is named as \emph{P\_Q} when feature P causes the vulnerability that enables exploit Q.

\subsection{Structure and Content}
\label{sec:repo-structure}

The repository contains top-level folders corresponding to various categories of vulnerabilities: \emph{ICC, Storage, System,} and \emph{Web}.  We refer to these top-level folders as \emph{category folders}.  Each category folder contains subfolders corresponding to different benchmarks.  We refer to these subfolders as \emph{benchmark folders}.  Each category folder also contains a README file that briefly describes each benchmark in the category.

There is one-to-one correspondence between benchmark folders and benchmarks.  Each benchmark folder is named as \emph{P\_Q} where P is the specific feature that causes a vulnerability of interest and Q is the exploit enabled by the vulnerability.  Each benchmark folder contains two \emph{app folders}: \emph{Benign} folder contains the source code of an app that uses feature P to exhibit a vulnerability and \emph{Malicious} folder contains the source code of an app that exploits the vulnerability exhibited by the app in the \emph{Benign} folder.  A README file in each benchmark folder summarizes the benchmark, describes the contained vulnerability and the corresponding exploit, provides instructions to build both Benign and Malicious apps (refer to \Fref{sec:repo-work-flow}), and lists the versions of Android on which the benchmark has been tested.

In case of \emph{Web} category, benchmark folders do not contain a \emph{Malicious} folder in them because the captured vulnerabilities can be exploited by Man-in-the-Middle (MitM) attacks.  This requires a web server that the Benign apps can connect to.  Consequently, code and instructions to set up local web server are provided in a top-level folder named \emph{Misc/LocalServer}.  README file of each Benign app contain instructions to configure the app to talk to the local web server.  As for the MitM attack in this set up, the users are free to choose how to mount such an attack.

Currently, the repository contains 25 benchmarks, each capturing a unique vulnerability.  There are 13 ICC benchmarks, 2 Storage benchmarks, 4 System benchmarks, and 6 Web benchmarks.  The smallest and the largest benchmarks contains 490 and 1510 lines of code and configuration, respectively.  In terms of bytes, Storage benchmarks are larger as they rely on two external APK files of size 13,144 KB in total.  \Fref{tab:repo-size-stats} provides basic size based statistics about the benchmarks. 

\begin{table*}[ht]
\centering
\small
\begin{tabular}{|c|r|r|r|r|r|r|r|r|r|}
  \cline{2-10}
  \multicolumn{1}{c}{} & \multicolumn{3}{|c|}{\textbf{Total Size (KB)}} & \multicolumn{3}{c|}{\textbf{Size of Text Files (KB)}} & \multicolumn{3}{c|}{\textbf{\# Lines of Code+Config}} \\
  \hline
  \textbf{Category} & \emph{Min} & \emph{Median} & {Max} & \emph{Min} & \emph{Median} & {Max} & \emph{Min} & \emph{Median} & {Max} \\
  \hline
  \textit{ICC} & 364 & 380 & 700 & 164 & 196 & 336 & 635 & 1182 & 1510 \\
  \hline
  \textit{Storage} & 1,000 & 7,500 & 14,000 & 156 & 160 & 164 & 936 & 955 & 974 \\
  \hline 
  \textit{System} & 364 & 404 & 444 & 164 & 172 & 180 & 968 & 995.5 & 1035 \\
  \hline 
  \textit{Web} & 192 & 224 & 228 & 84 & 86 & 100 & 490 & 522 & 590 \\
  \hline
\end{tabular}
\caption{Basic Source Code Size Statistics of Benchmarks in Ghera}
\label{tab:repo-size-stats}
\end{table*}

\subsection{Work Flow}
\label{sec:repo-work-flow}

To illustrate the steps involved in using Ghera, we consider the \textit{ICC/DynamicallyBroadcastReceiverRegistration-UnrestrictedAccess} \newline benchmark.  In this benchmark, the benign app dynamically registers a service with Android platform.  This action exposes the service to any app on the same device, including malicious apps.  Following are the instructions to use the benchmark to reproduce and exploit this vulnerability.

\begin{enumerate}
\item Execute the following commands to create an Android Virtual Device (AVD) if an AVD does not already exist.

\begin{minted}[linenos,fontsize=\small]{bash}
avdmanager list target
avdmanager list avd
avdmanager create avd -n <name> -k <target>
\end{minted}

\texttt{<name>} should be an identifier not in the list of existing AVD names displayed by the command in line 2. \texttt{<target>} should be a target identifier of a system image listed by the command in line 1.

\item Start the emulator.

\begin{minted}[fontsize=\small]{bash}
emulator -avd <name>
\end{minted}

\item Build Benign and Malicious apps and install them on to the emulator using the following commands.
\begin{minted}[fontsize=\small]{bash}
cd Benign
./gradlew installDebug
cd ../Malicious
./gradlew installDebug
\end{minted}

\item In the emulator, launch Benign app followed by Malicious app.

\item Execute the following command.
\begin{minted}[fontsize=\small]{bash}
adb logcat -d | grep "UserLeftBroadcastRecv"
\end{minted}
If the exploit was successful, you should see a message "An email will be sent to rookie@malicious.com with the text: I can send email without any permissions".
\end{enumerate}

The number of steps are few and they are simple.  However, they assume the user is familiar with Android development in terms of various tools (\eg virtual devices, emulators) used while developing Android apps.

Further, instructions assume users will use these benchmarks on emulators.  If a user wants to use a benchmark on a device, then she should skip step 2 and ensure an Android device is connected to the development machine and configured as a device.

The workflow for every benchmark is similar to the above workflow barring minor changes associated with specific aspects of the benchmark, if any.

\subsection{What about characteristics?}
\label{sec:repo-characteristics}

While we identified the benchmark and repository characteristics in retrospective, we believe it did influence the creation of Ghera and its benchmarks.  So, we will now examine if and why Ghera and its benchmarks exhibits the identified characteristics.

The creation of each benchmark in Ghera focused on the specific features of Android framework and how it could lead to a vulnerability.  Based on this information, we created a Benign app that exhibited the vulnerability.  We tested the vulnerability in the Benign app by using the Malicious app that exploited the vulnerability in the Benign app.  This process did not involve the use of any vulnerability (or exploit) detection tools to confirm the presence of the vulnerability.  Hence, the benchmarks are \emph{tool and technique agnostic}.

As described above, for each benchmark, we tested the presence of vulnerability in the Benign app by exploiting the vulnerability via the Malicious app.  As for Web apps, we used a local web server with self-signed certificates and static HTML containing malicious JavaScript code to exploit and verify the presence of the vulnerability.  Hence, the benchmarks are \emph{authentic}. 

When creating the Benign app of each benchmark capturing a vulnerability X due to features F, we only used features F to create X.  Further, we made very little use of other features of Android framework in the Benign app.  Hence, we claim the benchmarks are \emph{feature specific}.

The benchmarks focus on reproducing vulnerabilities while being minimal in size and the features used from Android framework.  This observation is partially supported by the basic source code size statistics of the benchmarks in \Fref{tab:repo-size-stats}.  Hence, we claim the benchmarks are \emph{contextual}.

Every Ghera benchmarks come with instructions to build, install, and execute its Benign and Malicious apps to exercise captured vulnerabilities.  Also, we have verified the instructions when testing the benchmarks on different versions of Android.  Further, the work flow associated with each benchmark as illustrated in \Fref{sec:repo-work-flow} is short, mostly automated, simple, and easy.  Hence, the benchmarks are both \emph{ready-to-use} and \emph{easy-to-use}.

We tested each benchmark on different supported versions of Android on emulators.  Based on the success of reproducing the vulnerabilities in these tests, we have documented the versions of Android on which a benchmark reproduces the captured vulnerability.  Hence, the benchmarks are \emph{version specific}.  Further, the benchmarks cover all of the supported versions of Android.

Each benchmark is accompanied by documentation that describes the vulnerability and the associated exploit along with instructions to reproduce and exploit the vulnerability.  In addition, each benchmark is available in source form.  Hence, the benchmarks are \emph{well documented}.

Since each benchmark (where possible) is composed of two apps: one exhibiting the vulnerability and another exploiting the exhibited vulnerability, the benchmarks exhibit the \emph{dual} characteristic.

Ghera is hosted as a public repository that accepts contribution from the community.  Hence, it is an \emph{open} repository.  

While Ghera does have benchmarks covering four different areas (capabilities) of Android framework, there are many more areas of the Android framework that may be associated with known vulnerabilities and are not covered by Ghera benchmarks.  Hence, Ghera is \emph{not yet a comprehensive} repository.

\subsection{Limitations \& Threats to Validity}
\label{sec:repo-limitations}

Currently, Ghera only caters lean benchmarks. Consequently, it cannot be used to evaluate the scalability of vulnerability detection tools, which would require fat benchmarks.

Ghera benchmarks currently capture 25 known vulnerabilities while covering four areas of Android framework API: ICC, Storage, System, and Web.  However, sources such as CVE \cite{CVE:URL} suggest vulnerabilities exists in other areas of the Android framework (\eg Networking, Camera) not covered by Ghera.  Likewise, in the four areas covered by Ghera, there may be known vulnerabilities that are not captured by any Ghera benchmarks.

In the previous section, we claimed Ghera benchmarks were feature specific and contextual based on our diligent process of creating benchmarks.  However, our claim does not account for any bias or oversight.

\section{Future Work}
\label{sec:future-work}

Here are few possibilities to extend and use Ghera to help secure Android apps.

\begin{enumerate}
  \item Add new benchmarks with vulnerabilities stemming from ICC, System, Storage, and Web related Android framework API but not present in existing benchmarks.  Also, add new benchmarks with vulnerabilities stemming from Android APIs such as Networking and Camera that have not been considered by current benchmarks.  These additions will make the repository more comprehensive.
  \item Extend the repository with real world apps (or links to apps in repositories such as AndroZoo) that have vulnerabilities captured by existing benchmarks as new context-specific benchmarks.  These additions will help evaluate techniques for scale.  
  \item Extract code patterns from these benchmarks to identify other instances of these benchmarks.  These patterns can be codified as IDE plugins to help developers avoid vulnerabilities while coding apps.  Also, these patterns could be used to measure the prevalence of corresponding vulnerabilities in real-world apps.
  \item Create benchmarks based on Android app vulnerabilities listed on CVE \cite{CVE:URL}.  These additions will help developers understand the reported vulnerabilities and explore solutions to avoid them.
  \item Use these benchmarks to evaluate existing vulnerability analysis tools.  Such an evaluation will help compare existing tools and possibly provide hints to new possibilities such as combining existing techniques to improve detection accuracy.  (We are currently pursuing such an evaluation of existing tools.)

\end{enumerate}

\section{Summary}
\label{sec:summary}

While Android security has been the focus of research efforts in the past few years, there are hardly any benchmarks to test and evaluate vulnerability detection tools that help develop secure Android apps.  Our search for such benchmarks led us to create Ghera, an open repository of vulnerability benchmarks (\url{http://bitbucket.org/secure-it-i/android-app-vulnerability-benchmarks}).  Currently, Ghera captures 25 known vulnerabilities in Android apps spanning four different capabilities of Android platform.  We plan to extend it and use it in an ongoing tools evaluation effort.  We hope the community will use and contribute to Ghera to improve the evaluation of Android app vulnerability detection tools and techniques. 

During the creation of Ghera, we uncovered various desirable characteristics of vulnerability benchmarks and benchmark repositories.  Given the increasing interest in rigorous empirical evaluation, we have documented these characteristics in the hope that it can serve as guidance to create benchmarks to assist with rigorous and reproducible evaluations.

\bibliographystyle{ACM-Reference-Format}
\bibliography{references} 


\begin{thebibliography}{00}


\ifx \showCODEN    \undefined \def \showCODEN     #1{\unskip}     \fi
\ifx \showDOI      \undefined \def \showDOI       #1{#1}\fi
\ifx \showISBNx    \undefined \def \showISBNx     #1{\unskip}     \fi
\ifx \showISBNxiii \undefined \def \showISBNxiii  #1{\unskip}     \fi
\ifx \showISSN     \undefined \def \showISSN      #1{\unskip}     \fi
\ifx \showLCCN     \undefined \def \showLCCN      #1{\unskip}     \fi
\ifx \shownote     \undefined \def \shownote      #1{#1}          \fi
\ifx \showarticletitle \undefined \def \showarticletitle #1{#1}   \fi
\ifx \showURL      \undefined \def \showURL       {\relax}        \fi
\providecommand\bibfield[2]{#2}
\providecommand\bibinfo[2]{#2}
\providecommand\natexlab[1]{#1}
\providecommand\showeprint[2][]{arXiv:#2}

\bibitem[\protect\citeauthoryear{Allix, Bissyande, Klein, and Traon}{Allix
  et~al\mbox{.}}{2016}]%
        {Allix:MSR16}
\bibfield{author}{\bibinfo{person}{Kevin Allix},
  \bibinfo{person}{Tegaw\'{e}nd\'{e}~F. Bissyande}, \bibinfo{person}{Jacques
  Klein}, {and} \bibinfo{person}{Yves~Le Traon}.}
  \bibinfo{year}{2016}\natexlab{}.
\newblock \showarticletitle{AndroZoo: Collecting Millions of Android Apps for
  the Research Community}. In \bibinfo{booktitle}{{\em Proceedings of the 13th
  International Conference on Mining Software Repositories}}.
  \bibinfo{publisher}{ACM}, \bibinfo{pages}{468--471}.
\newblock


\bibitem[\protect\citeauthoryear{Arzt, Rasthofer, Fritz, Bodden, Bartel, Klein,
  Le~Traon, Octeau, and McDaniel}{Arzt et~al\mbox{.}}{2014}]%
        {Arzt:PLDI14}
\bibfield{author}{\bibinfo{person}{Steven Arzt}, \bibinfo{person}{Siegfried
  Rasthofer}, \bibinfo{person}{Christian Fritz}, \bibinfo{person}{Eric Bodden},
  \bibinfo{person}{Alexandre Bartel}, \bibinfo{person}{Jacques Klein},
  \bibinfo{person}{Yves Le~Traon}, \bibinfo{person}{Damien Octeau}, {and}
  \bibinfo{person}{Patrick McDaniel}.} \bibinfo{year}{2014}\natexlab{}.
\newblock \showarticletitle{FlowDroid: Precise Context, Flow, Field,
  Object-sensitive and Lifecycle-aware Taint Analysis for Android Apps}. In
  \bibinfo{booktitle}{{\em Proceedings of the 35th ACM SIGPLAN Conference on
  Programming Language Design and Implementation}}. \bibinfo{publisher}{ACM},
  \bibinfo{pages}{259--269}.
\newblock
\showISBNx{978-1-4503-2784-8}


\bibitem[\protect\citeauthoryear{Bhattacharya}{Bhattacharya}{2016}]%
        {MarketShare1:URL}
\bibfield{author}{\bibinfo{person}{Ananya Bhattacharya}.}
  \bibinfo{year}{2016}\natexlab{}.
\newblock \bibinfo{title}{Android just hit a record 88\% market share of all
  smartphones}.
\newblock   (\bibinfo{year}{2016}).
\newblock
\showURL{%
\url{https://qz.com/826672/android-goog-just-hit-a-record-88-market-share-of-all-smartphones/}}
\newblock
\shownote{Accessed: 07-Jun-2017.}


\bibitem[\protect\citeauthoryear{Cao, Fratantonio, Bianchi, Egele, land
  Giovanni~Vigna, and Chen}{Cao et~al\mbox{.}}{2015}]%
        {Cao:NDSS15}
\bibfield{author}{\bibinfo{person}{Yinzhi Cao}, \bibinfo{person}{Yanick
  Fratantonio}, \bibinfo{person}{Antonio Bianchi}, \bibinfo{person}{Manuel
  Egele}, \bibinfo{person}{Christopher~Kruege land Giovanni~Vigna}, {and}
  \bibinfo{person}{Yan Chen}.} \bibinfo{year}{2015}\natexlab{}.
\newblock \showarticletitle{EdgeMiner: Automatically Detecting Implicit Control
  Flow Transitions through the Android Framework}. In \bibinfo{booktitle}{{\em
  Proceedings of the Network and Distributed System Security Symposium}}.
\newblock


\bibitem[\protect\citeauthoryear{Chin, Felt, Greenwood, and Wagner}{Chin
  et~al\mbox{.}}{2011}]%
        {Chin:Mobisys11}
\bibfield{author}{\bibinfo{person}{Erika Chin},
  \bibinfo{person}{Adrienne~Porter Felt}, \bibinfo{person}{Kate Greenwood},
  {and} \bibinfo{person}{David Wagner}.} \bibinfo{year}{2011}\natexlab{}.
\newblock \showarticletitle{Analyzing Inter-application Communication in
  {A}ndroid}. In \bibinfo{booktitle}{{\em Proceedings of the 9th International
  Conference on Mobile Systems, Applications, and Services}}.
  \bibinfo{publisher}{ACM}, \bibinfo{pages}{239--252}.
\newblock


\bibitem[\protect\citeauthoryear{Corporation}{Corporation}{2017}]%
        {CVE:URL}
\bibfield{author}{\bibinfo{person}{Mitre Corporation}.}
  \bibinfo{year}{2017}\natexlab{}.
\newblock \bibinfo{title}{Common Vulnerabilities and Exposures}.
\newblock   (\bibinfo{year}{2017}).
\newblock
\showURL{%
\url{https://cve.mitre.org/cgi-bin/cvekey.cgi?keyword=Android}}
\newblock
\shownote{Accessed: 08-Jun-2017.}


\bibitem[\protect\citeauthoryear{Elenkov}{Elenkov}{2015}]%
        {Elenkov:Book}
\bibfield{author}{\bibinfo{person}{Nikolay Elenkov}.}
  \bibinfo{year}{2015}\natexlab{}.
\newblock \bibinfo{booktitle}{{\em Android Security Internals}}.
\newblock \bibinfo{publisher}{No Starch Press}.
\newblock


\bibitem[\protect\citeauthoryear{Fahl, Harbach, Muders, Baumg\"{a}rtner,
  Freisleben, and Smith}{Fahl et~al\mbox{.}}{2012}]%
        {Fahl:CCS12}
\bibfield{author}{\bibinfo{person}{Sascha Fahl}, \bibinfo{person}{Marian
  Harbach}, \bibinfo{person}{Thomas Muders}, \bibinfo{person}{Lars
  Baumg\"{a}rtner}, \bibinfo{person}{Bernd Freisleben}, {and}
  \bibinfo{person}{Matthew Smith}.} \bibinfo{year}{2012}\natexlab{}.
\newblock \showarticletitle{Why Eve and Mallory Love Android: An Analysis of
  Android SSL (in)Security}. In \bibinfo{booktitle}{{\em Proceedings of the
  2012 ACM Conference on Computer and Communications Security}}.
  \bibinfo{publisher}{ACM}, \bibinfo{pages}{50--61}.
\newblock
\showISBNx{978-1-4503-1651-4}


\bibitem[\protect\citeauthoryear{Inc.}{Inc.}{2017}]%
        {AndroidSecTips:URL}
\bibfield{author}{\bibinfo{person}{Google Inc.}}
  \bibinfo{year}{2017}\natexlab{}.
\newblock \bibinfo{title}{Android Security Tips}.
\newblock
  \bibinfo{howpublished}{https://developer.android.com/training/articles/security-tips.html}.
    (\bibinfo{year}{2017}).
\newblock
\newblock
\shownote{Accessed: 07-Jun-2017.}


\bibitem[\protect\citeauthoryear{Inc. and Alliance}{Inc. and Alliance}{2017a}]%
        {AndroidSecurityBestPractices:URL}
\bibfield{author}{\bibinfo{person}{Google Inc.} {and}
  \bibinfo{person}{Open~Handset Alliance}.} \bibinfo{year}{2017}\natexlab{a}.
\newblock \bibinfo{title}{Best Practices for Security \& Privacy}.
\newblock   (\bibinfo{year}{2017}).
\newblock
\showURL{%
\url{https://developer.android.com/training/best-security.html}}
\newblock
\shownote{Accessed: 08-Jun-2017.}


\bibitem[\protect\citeauthoryear{Inc. and Alliance}{Inc. and Alliance}{2017b}]%
        {AndroidSecurity:URL}
\bibfield{author}{\bibinfo{person}{Google Inc.} {and}
  \bibinfo{person}{Open~Handset Alliance}.} \bibinfo{year}{2017}\natexlab{b}.
\newblock \bibinfo{title}{Security}.
\newblock   (\bibinfo{year}{2017}).
\newblock
\showURL{%
\url{https://source.android.com/security/overview/implement}}
\newblock
\shownote{Accessed: 08-Jun-2017.}


\bibitem[\protect\citeauthoryear{(JSSEC)}{(JSSEC)}{2016}]%
        {JSSEC:Book}
\bibfield{author}{\bibinfo{person}{Japan Smartphone Security~Association
  (JSSEC)}.} \bibinfo{year}{2016}\natexlab{}.
\newblock \bibinfo{booktitle}{{\em Android Application Secure Design/Secure
  Coding Guidebook}}.
\newblock


\bibitem[\protect\citeauthoryear{Luo, Hao, Du, Wang, and Yin}{Luo
  et~al\mbox{.}}{2011}]%
        {Luo:ACSAC11}
\bibfield{author}{\bibinfo{person}{Tongbo Luo}, \bibinfo{person}{Hao Hao},
  \bibinfo{person}{Wenliang Du}, \bibinfo{person}{Yifei Wang}, {and}
  \bibinfo{person}{Heng Yin}.} \bibinfo{year}{2011}\natexlab{}.
\newblock \showarticletitle{Attacks on WebView in the Android system}. In
  \bibinfo{booktitle}{{\em Proceedings of the 27th Annual Computer Security
  Applications Conference}}. ACM, \bibinfo{pages}{343--352}.
\newblock


\bibitem[\protect\citeauthoryear{Oberheide and Miller}{Oberheide and
  Miller}{2012}]%
        {Bouncer:URL}
\bibfield{author}{\bibinfo{person}{Oberheide} {and} \bibinfo{person}{Miller}.}
  \bibinfo{year}{2012}\natexlab{}.
\newblock \bibinfo{title}{Dissecting {G}oogle {B}ouncer}.
\newblock
  \bibinfo{howpublished}{https://jon.oberheide.org/files/summercon12-bouncer.pdf}.
    (\bibinfo{year}{2012}).
\newblock
\newblock
\shownote{Accessed: 08-Jun-2017.}


\bibitem[\protect\citeauthoryear{Octeau, McDaniel, Jha, Bartel, Bodden, Klein,
  and Le~Traon}{Octeau et~al\mbox{.}}{2013}]%
        {Octeau:SEC13}
\bibfield{author}{\bibinfo{person}{Damien Octeau}, \bibinfo{person}{Patrick
  McDaniel}, \bibinfo{person}{Somesh Jha}, \bibinfo{person}{Alexandre Bartel},
  \bibinfo{person}{Eric Bodden}, \bibinfo{person}{Jacques Klein}, {and}
  \bibinfo{person}{Yves Le~Traon}.} \bibinfo{year}{2013}\natexlab{}.
\newblock \showarticletitle{Effective Inter-component Communication Mapping in
  Android with Epicc: An Essential Step Towards Holistic Security Analysis}. In
  \bibinfo{booktitle}{{\em Proceedings of the 22Nd USENIX Conference on
  Security}}. \bibinfo{publisher}{USENIX Association},
  \bibinfo{pages}{543--558}.
\newblock
\showISBNx{978-1-931971-03-4}


\bibitem[\protect\citeauthoryear{Reaves, Bowers, Gorski~III, Anise, Bobhate,
  Cho, Das, Hussain, Karachiwala, Scaife, Wright, Butler, Enck, and
  Traynor}{Reaves et~al\mbox{.}}{2016}]%
        {Reaves:CSUR16}
\bibfield{author}{\bibinfo{person}{Bradley Reaves}, \bibinfo{person}{Jasmine
  Bowers}, \bibinfo{person}{Sigmund~Albert Gorski~III},
  \bibinfo{person}{Olabode Anise}, \bibinfo{person}{Rahul Bobhate},
  \bibinfo{person}{Raymond Cho}, \bibinfo{person}{Hiranava Das},
  \bibinfo{person}{Sharique Hussain}, \bibinfo{person}{Hamza Karachiwala},
  \bibinfo{person}{Nolen Scaife}, \bibinfo{person}{Byron Wright},
  \bibinfo{person}{Kevin Butler}, \bibinfo{person}{William Enck}, {and}
  \bibinfo{person}{Patrick Traynor}.} \bibinfo{year}{2016}\natexlab{}.
\newblock \showarticletitle{*Droid: Assessment and Evaluation of Android
  Application Analysis Tools}.
\newblock \bibinfo{journal}{{\em ACM Comput. Surv.\/}} \bibinfo{volume}{49},
  \bibinfo{number}{3}, Article \bibinfo{articleno}{55} (\bibinfo{date}{Oct.}
  \bibinfo{year}{2016}), \bibinfo{numpages}{30}~pages.
\newblock
\showISSN{0360-0300}


\bibitem[\protect\citeauthoryear{Reaves, Scaife, Bates, Traynor, and
  Butler}{Reaves et~al\mbox{.}}{2015}]%
        {Reaves:USENIX15}
\bibfield{author}{\bibinfo{person}{Bradley Reaves}, \bibinfo{person}{Nolen
  Scaife}, \bibinfo{person}{Adam Bates}, \bibinfo{person}{Patrick Traynor},
  {and} \bibinfo{person}{Kevin~R.B. Butler}.} \bibinfo{year}{2015}\natexlab{}.
\newblock \showarticletitle{Mo(bile) Money, Mo(bile) Problems: Analysis of
  Branchless Banking Applications in the Developing World}. In
  \bibinfo{booktitle}{{\em 24th USENIX Security Symposium}}.
  \bibinfo{publisher}{USENIX Association}, \bibinfo{pages}{17--32}.
\newblock


\bibitem[\protect\citeauthoryear{Rudrapp}{Rudrapp}{2015}]%
        {McAffee:Rudrapp15}
\bibfield{author}{\bibinfo{person}{Naveen Rudrapp}.}
  \bibinfo{year}{2015}\natexlab{}.
\newblock \bibinfo{title}{Secure Coding for {A}ndroid Applications}.
\newblock
  \bibinfo{howpublished}{http://www.mcafee.com/us/resources/white-papers/foundstone/wp-secure-coding-android-applications.pdf}.
    (\bibinfo{year}{2015}).
\newblock
\newblock
\shownote{Accessed: 08-Jun-2017.}


\bibitem[\protect\citeauthoryear{Sadeghi, Bagheri, Garcia, and Malek}{Sadeghi
  et~al\mbox{.}}{2016}]%
        {Sadeghi:TR2016}
\bibfield{author}{\bibinfo{person}{Alireza Sadeghi}, \bibinfo{person}{Hamid
  Bagheri}, \bibinfo{person}{Joshua Garcia}, {and} \bibinfo{person}{Sam
  Malek}.} \bibinfo{year}{2016}\natexlab{}.
\newblock \bibinfo{booktitle}{{\em A Taxonomy and Qualitative Comparison of
  Program Analysis Techniques for Security Assessment of Android Apps}}.
\newblock \bibinfo{type}{{T}echnical {R}eport}.
  \bibinfo{institution}{University of California, Irvine}.
\newblock


\bibitem[\protect\citeauthoryear{Sadeghi, Bagheri, and Malek}{Sadeghi
  et~al\mbox{.}}{2015}]%
        {Sadeghi:ICSE15}
\bibfield{author}{\bibinfo{person}{A. Sadeghi}, \bibinfo{person}{H. Bagheri},
  {and} \bibinfo{person}{S. Malek}.} \bibinfo{year}{2015}\natexlab{}.
\newblock \showarticletitle{Analysis of Android Inter-App Security
  Vulnerabilities Using COVERT}. In \bibinfo{booktitle}{{\em 2015 IEEE/ACM 37th
  IEEE International Conference on Software Engineering}}.
  \bibinfo{pages}{725--728}.
\newblock


\bibitem[\protect\citeauthoryear{SEI}{SEI}{2016}]%
        {SEIRules:URL}
\bibfield{author}{\bibinfo{person}{SEI}.} \bibinfo{year}{2016}\natexlab{}.
\newblock \bibinfo{title}{Android Secure Coding Standard}.
\newblock   (\bibinfo{year}{2016}).
\newblock
\showURL{%
\url{https://www.securecoding.cert.org/confluence/pages/viewpage.action?pageId=111509535}}
\newblock
\shownote{Accessed: 17-Jun-2017.}


\bibitem[\protect\citeauthoryear{Statista}{Statista}{2017}]%
        {GooglePlayAppCount:URL}
\bibfield{author}{\bibinfo{person}{Statista}.} \bibinfo{year}{2017}\natexlab{}.
\newblock \bibinfo{title}{Number of available applications in the Google Play
  Store from December 2009 to March 2017}.
\newblock   (\bibinfo{year}{2017}).
\newblock
\showURL{%
\url{https://www.statista.com/statistics/266210/number-of-available-applications-in-the-google-play-store/}}
\newblock
\shownote{Accessed: 08-Jun-2017.}


\bibitem[\protect\citeauthoryear{Sufatrio, Tan, Chua, and Thing}{Sufatrio
  et~al\mbox{.}}{2015}]%
        {Sufatrio:CSUR15}
\bibfield{author}{\bibinfo{person}{Sufatrio}, \bibinfo{person}{Darell J.~J.
  Tan}, \bibinfo{person}{Tong-Wei Chua}, {and} \bibinfo{person}{Vrizlynn L.~L.
  Thing}.} \bibinfo{year}{2015}\natexlab{}.
\newblock \showarticletitle{Securing {A}ndroid: A Survey, Taxonomy, and
  Challenges}.
\newblock \bibinfo{journal}{{\em ACM Comput. Surv.\/}} \bibinfo{volume}{47},
  \bibinfo{number}{4} (\bibinfo{year}{2015}), \bibinfo{pages}{58:1--58:45}.
\newblock


\bibitem[\protect\citeauthoryear{Viennot, Garcia, and Nieh}{Viennot
  et~al\mbox{.}}{2014}]%
        {Viennot:SIGMETRICS14}
\bibfield{author}{\bibinfo{person}{Nicolas Viennot}, \bibinfo{person}{Edward
  Garcia}, {and} \bibinfo{person}{Jason Nieh}.}
  \bibinfo{year}{2014}\natexlab{}.
\newblock \showarticletitle{A Measurement Study of Google Play}. In
  \bibinfo{booktitle}{{\em The 2014 ACM International Conference on Measurement
  and Modeling of Computer Systems}}. \bibinfo{publisher}{ACM},
  \bibinfo{pages}{221--233}.
\newblock


\bibitem[\protect\citeauthoryear{Wei, Roy, Ou, and Robby}{Wei
  et~al\mbox{.}}{2014}]%
        {Wei:CCS14}
\bibfield{author}{\bibinfo{person}{Fengguo Wei}, \bibinfo{person}{Sankardas
  Roy}, \bibinfo{person}{Xinming Ou}, {and} \bibinfo{person}{Robby}.}
  \bibinfo{year}{2014}\natexlab{}.
\newblock \showarticletitle{Amandroid: A Precise and General Inter-component
  Data Flow Analysis Framework for Security Vetting of Android Apps}. In
  \bibinfo{booktitle}{{\em Proceedings of the 2014 ACM SIGSAC Conference on
  Computer and Communications Security}}. \bibinfo{publisher}{ACM},
  \bibinfo{pages}{1329--1341}.
\newblock
\showISBNx{978-1-4503-2957-6}


\bibitem[\protect\citeauthoryear{Zhou and Jiang}{Zhou and Jiang}{2012}]%
        {Zhou:IEEESSP12}
\bibfield{author}{\bibinfo{person}{Y. Zhou} {and} \bibinfo{person}{X. Jiang}.}
  \bibinfo{year}{2012}\natexlab{}.
\newblock \showarticletitle{Dissecting Android Malware: Characterization and
  Evolution}. In \bibinfo{booktitle}{{\em 2012 IEEE Symposium on Security and
  Privacy}}. \bibinfo{pages}{95--109}.
\newblock


\bibitem[\protect\citeauthoryear{Zhou, Wang, Zhou, and Jiang}{Zhou
  et~al\mbox{.}}{2012}]%
        {Zhou:NDSS12}
\bibfield{author}{\bibinfo{person}{Yajin Zhou}, \bibinfo{person}{Zhi Wang},
  \bibinfo{person}{Wu Zhou}, {and} \bibinfo{person}{Xuxian Jiang}.}
  \bibinfo{year}{2012}\natexlab{}.
\newblock \showarticletitle{Hey, You, Get Off of My Market: Detecting Malicious
  Apps in Official and Alternative {A}ndroid Markets}. In
  \bibinfo{booktitle}{{\em 19th Annual Network and Distributed System Security
  Symposium, {NDSS} 2012, San Diego, California, USA, February 5-8, 2012}}.
\newblock


\end{thebibliography}

\appendix
\section{Catalog of Benchmarks}
\label{sec:catalog}

In this section, we catalog the current benchmarks in Ghera according to the vulnerability categories identified in \Fref{sec:repo-design-choices}.  For each benchmark, we provide a short description of the vulnerability and the exploit that uses the vulnerability.

\subsection{Inter Component Communication}
\label{sec:cat-icc}

Android apps are composed of four basic kinds of components: 1) \textit{Activity} components display the user interface, 2) \textit{Service} components perform background operations, 3) \textit{Broadcast Receiver} components receive event notifications and act on those notifications, and 4) \textit{Content Provider} components manage app data.  Communication between components in an app and in different apps is facilitated via exchange of \textit{Intent}s.  Components specify their ability to process specific kinds of intents by using \emph{intent-filters}.

\subsubsection{Dynamically registered broadcast receiver provides unrestricted access}

\paragraph{Vulnerability:} When a broadcast receiver is dynamically registered with the Android platform, a non-null intent filter is provided.  As a result, the component is automatically exported to be accessible from other apps, including malicious apps.

\paragraph{Exploit:} A malicious app broadcasts a message to a dynamically registered broadcast receiver.  This triggers the broadcast receiver to process the intent and unintentionally perform an action on behalf of the malicious app.

\subsubsection{Empty pending intent leaks privilege}

\paragraph{Vulnerability:} An app X can allow another app Y to perform an action on its behalf at a future time via a \textit{pending intent}; these intents are saved in the system. When no action is specified in a pending intent, the recipient of the pending intent can set any action and execute it in the context of the app that sent the pending intent.

\paragraph{Exploit:} A malicious app specifies its interest in the pending intent via an intent-filter. Upon receiving an empty pending intent, the malicious app associates a malicious action with the pending intent. Conseuqently, when the pending intent is processed, the malicious action will be executed in the context of app X.

\subsubsection{Low priority activity prone to hijacking}

\paragraph{Vulnerability:} A \emph{priority} can be specified for an activity in the app's manifest file.  When an activity is started, Android displays all activities with the same intent-filter as a list to the user in the order of priority (high to low).

\paragraph{Exploit:} A malicious app registers an activity X with the same \textit{intent-filter} as that of an activity Y registered by a benign app and with higher priority than Y. Consequently, the malicious app's activity X will be displayed before the benign app's activity Y.

\subsubsection{Service started by implicit intent is prone to hijacking}

\paragraph{Vulnerability:} Android platform uses intent-filters to identify the service to process \emph{implicit intents}, \ie intents dedicated to a class of targets (as opposed to specific target).  When multiple services have the same intent-filter, the service with higher priority is chosen to process corresponding intents.

\paragraph{Exploit:} A malicious app has a service X with the same intent-filter as that of the service Y in a benign app and with higher priority than Y.  When an app requests the start of service Y by relying on the intent-filter, service X in the malicious app will be started.

\subsubsection{Implicit pending intent leaks information}

\paragraph{Vulnerability:} A app X can create a pending intent containing an implicit intent.  When the pending intent is processed, the containing implicit intent will be processed by a component identified based on the intent-filter.  When multiple components have the same intent-filter, the component with higher priority is chosen to process corresponding intents.

\paragraph{Exploit:} A malicious app has a component X with an intent-filter same as that of the component Y in the benign app and X has higher priority than Y.  So, component X is chosen (over component Y) to process the implicit intent in the pending intent.

\subsubsection{Content provider with inadequate path-permission leaks information}

\paragraph{Vulnerability:} An app can use \textit{path-permissions} to control access to the data exposed by a content provider. When an app protects a folder by permissions, only the files in the folder are protected by the permissions; none of the subfolders and their descendants are protected by the permissions.

\paragraph{Exploit:} A malicious app calls methods of a content provider to access and modify sub-directories and contained files that are not protected by path-permissions.

\subsubsection{Apps have unrestricted access to Broadcast receivers registered for system events}

\paragraph{Vulnerability:} When a Broadcast receiver registers to receive (system) intents from the Android platform, it needs to be exported. Consequently, it is accessible by any app without restrictions.

\paragraph{Exploit:} A malicious app sends an intent to a broadcast receiver that is registered to receive system intents and possibly forces it to perform unintended operations.

\subsubsection{Ordered broadcasts allow malicious data injection}

\paragraph{Vulnerability:} When an ordered broadcast is sent, broadcast receivers respond to it in the order of priority. Broadcast receivers with higher priority respond first and forward it to receivers with lower priority.

\paragraph{Exploit:} A malicious receiver with high priority receives the intent, changes it, and forwards it to lower priority receivers.

\subsubsection{Sticky broadcasts are prone to leaking sensitive information and malicious data injection}

\paragraph{Vulnerability:} When a \textit{sticky broadcast message (intent)} is sent, it is delivered to every registered receiver and is saved in the system to be provided to receivers that register for the message in the future.  When the message is re-broadcasted with modification, the modified message replaces the original message in the system.

\paragraph{Exploit:} A malicious broadcast receiver registers for the message at later time and retrieves any sensitive information in the message.  Further, it can modify the contents of the message and re-broadcast to provide incorrect information to future receivers of the message.

\subsubsection{Task affinity makes an app vulnerable to phishing attacks}

\paragraph{Vulnerability:} A \emph{task} is a collection (stack) of activities. When an activity is started, it is launched in a task.  An activity can request that it be started in a specific task.  This is known as \emph{task affinity}.  The task containing the displayed activity is moved to the background if none of the activities in that task are being displayed. When any activity from a task in the background is resumed, then the activity at the top of the task (and not the resumed activity) is displayed.  

\paragraph{Exploit:} An activity X in a malicious app requests to start itself in the same task as an activity Y in a benign app. When activity X is at the top of the task, any call to activity Y will cause activity X to be displayed to the user.

\subsubsection{Task affinity and task re-parenting enables phishing and denial-of-service}

\paragraph{Vulnerability:} An activity can request to always be at the top of a task.  This is called \textit{task re-parenting}.  In such cases, when an activity from that task resumed, activity at the top of the task will be displayed to the user.  

\paragraph{Exploit:} An activity in a malicious app uses task affinity and task re-parenting to supersede activities from other apps in a task and launch a denial-of-service attack or a phishing attack.

\subsubsection{Content Provider API allow unauthorized access}

\paragraph{Vulnerability:} Content provider API provides a method \texttt{call} to call any provider-defined method. With a reference to the content provider, this method can be invoked without any restrictions.

\paragraph{Exploit:} A malicious app uses \texttt{call} method to invoke content provider methods to access the underlying data even when it does not have specific permissions to access this data.

\subsection{Storage}
\label{sec:cat-storage}

Android provides numerous options for storing application data. It provides
\begin{enumerate}
\item \emph{Internal Storage} to store data that is private to apps. Every time an application is uninstalled, its  internal storage is emptied. Starting from Android 7.0, files stored in internal storage cannot be shared with other apps.
\item \emph{External Storage} as a data storage area that is common to apps.  Its public partition is accessible to any app without any restrictions.  Its private partition is only accessible to apps with a specific permission.
\end{enumerate}

\subsubsection{External storage allows data injection attack}

\paragraph{Vulnerability:} Files stored in external storage can be modified by an app with (appropriate) access to external storage.

\paragraph{Exploit:} A malicious app modifies external storage (\eg add files) and the content in external storage (\eg change files).

\subsubsection{Writing sensitive information to external storage enables information leak}

\paragraph{Vulnerability:} Files stored in external storage can be accessed by an app with (appropriate) access to external storage.

\paragraph{Exploit:} A malicious app reads content from external storage.

\subsection{System}
\label{sec:cat-system}
System APIs help Android apps access low level features of the Android platform like process management, thread management, runtime permissions etc. 

Every Android app runs in its own process with a unique Process ID (PID) and a User ID (UID).  All components in an app run in the same process.  A permission can be granted to an app at installation time or at run time.  If an app is granted a specific permission at installation time, then all components of the app are granted the same permission.  If component in an app is protected by a permission, only components that have been granted this permission can communicate with the protected component.  If the permission is checked at runtime, then all components have to request for the required permission.

\subsubsection{checkCallingOrSelfPermission method leaks privilege}

\paragraph{Vulnerability:} Before servicing a request, a component protected by a permission uses \texttt{checkCallingOrSelfPermission} to check if the requesting component has the permission. This method returns true if the app containing the requesting component or the app containing the protected component has the given permission.  When the app containing the protected component has the permission, the method will always return true. 

\paragraph{Exploit:} A malicious app accesses a component that is protected by permission P, is in an app that has permission P, and uses \texttt{checkCallingOrSelfPermission} to check for permission.

\subsubsection{checkPermission method leaks privilege}

\paragraph{Vulnerability:} Before servicing a request, a component protected by a permission uses \texttt{checkPermission} to check if the given PID and UID pair has the permission.  Typically, \texttt{getCallingPID} and \texttt{getCallingUID} methods of Binder API are used to retrieve PID and UID, respectively.  When these methods are invoked in the main thread of an app, they return the IDs of the app and not the IDs of the calling app.

\paragraph{Exploit:} A malicious app accesses a component that is protected by permission P, is in an app that has permission P, and uses \texttt{checkPermission} to check for permission in the main thread of the containing app.

\subsubsection{enforceCallingOrSelfPermission method leaks privilege}

\paragraph{Vulnerability:} Before servicing a request, a component protected by a permission uses \texttt{enforceCallingOrSelfPermission} to check if the requesting component has the permission. This method raises \texttt{SecurityException} if the app containing the requesting component or the app containing the protected component does not have the given permission.  When the app containing the protected component has the permission, the method will complete without any exceptions.

\paragraph{Exploit:} A malicious app accesses a component that is protected by permission P, is in an app that has permission P, and uses \texttt{enforceCallingOrSelfPermission} to enforce the permission.

\subsubsection{enforcePermission method leaks privilege}

\paragraph{Vulnerability:} Before servicing a request, a component protected by a permission uses \texttt{enforcePermission} to check if the given PID and UID pair has the permission.  Typically, \texttt{getCallingPID} and \texttt{getCallingUID} methods of Binder API are used to retrieve PID and UID, respectively.  When these methods are invoked in the main thread of an app, they return the IDs of the app and not the IDs of the calling app.

\paragraph{Exploit:} A malicious app accesses a component that is protected by permission P, is in an app that has permission P, and uses \texttt{enforcePermission} to enforce the permission in the main thread of the containing app.

\subsection{Web}
\label{sec:cat-web}

Web APIs allow Android apps to interact with web servers both insecurely and securely (via SSL/TLS), display web content through \textit{WebView} widget, and control navigation between web pages via \textit{WebViewClient} class.

\subsubsection{Incorrect hostname verification enables Man-in-the-Middle (MitM) attack}

\paragraph{Vulnerability:} Android apps that use SSL/TLS for secure communication employ custom implementations of the \texttt{HostnameVerifier} interface.  Such implementations perform custom checks on the given hostname in the \texttt{verify} method.  When these custom checks are incorrect or weak (\eg does not check hostname), then apps can end up connecting to malicious servers.

\paragraph{Exploit:} An application takes advantage of incorrect/weak hostname verification and mounts a MitM attack.

\subsubsection{Incorrect trust validation enables Man-in-the-Middle attack}

\paragraph{Vulnerability:} An app uses custom implementation of the \textit{TrustManager} interface to check if the presented certificates are valid and can be trusted. When these implementations are incorrect (\eg an empty stub implementation), unknown certificates may be trusted.  

\paragraph{Exploit:} An application takes advantage of incorrect trust validation and mounts a MitM attack.

\subsubsection{Allowing execution of unverified JavaScript code in WebView exposes app's resources}

\paragraph{Vulnerability:} When an app uses \texttt{WebView} to display web content and any JavaScript code embedded in the web content is executed, the code is executed with the same permission as the \texttt{WebView} instance used in the app.

\paragraph{Exploit:} An app injects malicious JavaScript code into the web content loaded in \texttt{WebView}, \eg modify static web page stored on the device.

\subsubsection{Ignoring SSL errors in WebViewClient enables Man-in-the-Middle (MitM) attack}

\paragraph{Vulnerability:} When an app loads web content from a SSL connection via \texttt{WebView} and is notified of an SSL error while loading the content (via \texttt{onReceivedSslError} method of \texttt{WebViewClient}), the app ignores the error.

\paragraph{Exploit:} An application takes advantage of ignored errors and mounts a MitM attack.

\subsubsection{Lack of validation of resource load requests in WebView allows loading malicious content}

\paragraph{Vulnerability:} When a resource (\eg CSS file, JavaScript file) is loaded in a web page in \texttt{WebView}, the app does not validate the resource load request in \texttt{shouldInterceptRequest} method of \texttt{WebViewClient}.  Consequently, any resource will be loaded into \texttt{WebView}.

\paragraph{Exploit:} An application takes advantage of lack of validation of resource load requests and mounts a MitM attack.

\subsubsection{Lack of validation web page load requests in WebView allows loading malicious content}

\begin{sloppypar}
\paragraph{Vulnerability:} When a web page is to be loaded into \texttt{WebView}, the app does not validate the web page load request in \texttt{shouldOverridUrlLoading} method of \texttt{WebViewClient}.  Consequently, any web page provided by the server will be loaded into \texttt{WebView}.
\end{sloppypar}

\paragraph{Exploit:} An application takes advantage of lack of validation of web page load requests and mounts a MitM attack.

\end{document}